\documentstyle[12pt,epsfig]{article}
                                                                                
\begin{document}                                                                
\date{}
                                                                                
\title{                                                                         
{\vspace{-1em} \normalsize                                                      
\hfill \parbox{50mm}{DESY 98-165}}\\[25mm]
Evidence for discrete chiral symmetry breaking               \\
in $N=1$ supersymmetric Yang-Mills theory                    \\[8mm]}
\author{DESY-M\"unster Collaboration                         \\[8mm]
R. Kirchner, I. Montvay, J. Westphalen                       \\
Deutsches Elektronen-Synchrotron DESY,                       \\
Notkestr.\,85, D-22603 Hamburg, Germany                      \\[5mm]
S. Luckmann, K. Spanderen                                    \\
Institut f\"ur Theoretische Physik I,                        \\
Universit\"at M\"unster, Wilhelm-Klemm-Str. 9,               \\
D-48149 M\"unster, Germany}
                                                                                
%%%%%%%%%%%%%%%%%%%%%%%%%%%%%%%%%%%%%%%%%%%%%%%%%%%%%%%%%%%%%%%%%%%%%%%%        
\newcommand{\be}{\begin{equation}}                                              
\newcommand{\ee}{\end{equation}}                                                
\newcommand{\half}{\frac{1}{2}}                                                 
\newcommand{\rar}{\rightarrow}                                                  
\newcommand{\lar}{\leftarrow}                                                   
%%%%%%%%%%%%%%%%%%%%%%%%%%%%%%%%%%%%%%%%%%%%%%%%%%%%%%%%%%%%%%%%%%%%%%%%
                                                                                
\maketitle
\vspace{3em}
                                                                                
\begin{abstract} \normalsize
 In a numerical Monte Carlo simulation of SU(2) Yang-Mills theory with
 dynamical gauginos we find evidence for two degenerate ground states
 at the supersymmetry point corresponding to zero gaugino mass.
 This is consistent with the expected pattern of spontaneous discrete
 chiral symmetry breaking $Z_4 \to Z_2$ caused by gaugino condensation.
\end{abstract}       

%%%%%%%%%%%%%%%%%%%%%%%%%%%%%%%%%%%%%%%%%%%%%%%%%%%%%%%%%%%%%%%%%%%%%%%%
\newpage
\section{Introduction}\label{sec1}
 The basic assumption about the non-perturbative dynamics of
 supersymmetric Yang-Mills (SYM) theory is that there is confinement
 and spontaneous chiral symmetry breaking, similar to QCD \cite{AKMRV}.
 (For a more recent introduction and review see also \cite{PESKIN}.)
 In the past years there has been great progress in the understanding
 of the non-perturbative properties of supersymmetric gauge theories,
 in particular following the seminal papers of Seiberg and Witten
 \cite{SEIWIT}.
 In case of $N=1$ SYM theory the non-perturbative results are not
 rigorous but fit into a self-consistent plausible picture of low
 energy dynamics of supersymmetric QCD (SQCD) \cite{SEIBERG}.
 The features of the low energy dynamics, like symmetries and bound
 state spectra, are formulated in terms of low energy effective actions
 \cite{VENYAN,FAGASCH}.
 Lattice Monte Carlo simulations may contribute by directly testing
 some of these predictions.

 The expected pattern of spontaneous chiral symmetry breaking in SYM
 theories is quite interesting: considering for definiteness the gauge
 group SU($N_c$), the expected symmetry breaking is $Z_{2N_c} \to Z_2$.
 This is because the global chiral symmetry of the gaugino (a Majorana
 fermion in the adjoint representation) is anomalous.
 The symmetry transformations are
\be\label{eq01}
\Psi_x \to e^{-i\varphi\gamma_5}\Psi_x \ , \hspace{2em}
\bar{\Psi}_x \to \bar{\Psi}_x e^{-i\varphi\gamma_5} \ ,
\ee
 where the Dirac-Majorana fields are used which satisfy, with the
 charge-conjugation Dirac matrix $C$,
\be\label{eq02}
\Psi_x = C \overline{\Psi}_x \ , \hspace{2em}
\overline{\Psi}_x = \Psi_x^T C \ .
\ee
 The group of symmetry transformations in (\ref{eq01}) coincide with
 the $R$-symmetry and hence will be called ${\rm U(1)}_R$.
 The transformation is equivalent to the transformation of the gaugino
 mass $m_{\tilde{g}}$ and a shift of the $\theta$-parameter:
\be\label{eq03}
m_{\tilde{g}} \to m_{\tilde{g}} e^{-2i\varphi\gamma_5} \ , \hspace{2em}
\theta \to \theta - 2N_c\varphi \ .
\ee
 Since $\theta$ is periodic with period $2\pi$, in the supersymmetric
 case with $m_{\tilde{g}}=0$ the ${\rm U(1)}_R$ symmetry is unbroken if
\be\label{eq04}
\varphi = \varphi_k \equiv \frac{k\pi}{N_c} \ , \hspace{2em}
(k=0,1,\ldots,2N_c-1) \ .
\ee
 Gaugino condensation means a non-zero vacuum expectation value
\be\label{eq05}
\langle \bar{\Psi}_x \Psi_x \rangle =
\langle \lambda_x^\alpha\lambda_{x\alpha} + 
\bar{\lambda}_x^{\dot{\alpha}} \bar{\lambda}_{x\dot{\alpha}} \rangle
\ne 0 \ .
\ee
 (Here, besides the Dirac-Majorana field, the Weyl-Majorana field
 components $\lambda_x^\alpha$ and $\bar{\lambda}_x^{\dot{\alpha}}$ are
 also introduced.)
 The gaugino condensate is transformed under ${\rm U(1)}_R$ according to
\be\label{eq06}
\langle \bar{\Psi}_x \Psi_x \rangle \to 
\langle \bar{\Psi}_x e^{-2i\varphi\gamma_5} \Psi_x \rangle \ .
\ee
 If such a condensate is produced by the dynamics then it breaks
 the $Z_{2N_c}$ symmetry to $Z_2$: the expected spontaneous chiral
 symmetry breaking is $Z_{2N_c} \to Z_2$.
 This implies the existence of $N_c$ discrete degenerate ground
 (vacuum) states with different orientations of the gaugino condensate
 according to (\ref{eq04}), (\ref{eq06}).

 A non-zero gaugino mass ($m_{\tilde{g}} \ne 0$) breaks the
 supersymmetry softly.
 As a function of the gaugino mass the degeneracy of the $N_c$ ground
 states is resolved.
 At $m_{\tilde{g}}=0$ the lowest ground state is changing.
 This gives rise to a characteristic pattern of first order phase
 transitions.

 In the special case of SU(2) gauge group, which will be considered in
 this paper, we have $Z_4 \to Z_2$ and in the two vacua the gaugino
 condensate has opposite signs.
 At $m_{\tilde{g}}=0$ the lowest ground states are exchanged and a first
 order phase transition occurs.
 In this letter we report on a large scale numerical Monte Carlo
 simulation with the aim to find numerical evidence for the existence of
 this phase transition. 

%%%%%%%%%%%%%%%%%%%%%%%%%%%%%%%%%%%%%%%%%%%%%%%%%%%%%%%%%%%%%%%%%%%%%%%%
\section{Lattice formulation}\label{sec2}
 The definition of an Euclidean path integral for Majorana fermions
 \cite{NICOLAI} may be obtained by starting from the well known
 Wilson formulation \cite{WILSON} of a Dirac fermion in the
 adjoint representation.
 If the Grassmanian fermion fields in the adjoint representation are
 denoted by $\psi^r_x$ and $\overline{\psi}^{\ r}_x$, with $r$ being the
 adjoint representation index, then
 the fermionic part of the lattice action can be written as
\be  \label{eq07}
S_f = \sum_{xu,yv} \overline{\psi}^{\ v}_y Q_{yv,xu} \psi^u_x \ .
\ee
 Here the {\em fermion matrix} $Q$ is defined by
\be  \label{eq08}
Q_{yv,xu} \equiv Q_{yv,xu}[U] \equiv
\delta_{yx}\delta_{vu} - K \sum_{\mu=1}^4 \left[
\delta_{y,x+\hat{\mu}}(1+\gamma_\mu) V_{vu,x\mu} +
\delta_{y+\hat{\mu},x}(1-\gamma_\mu) V^T_{vu,y\mu} \right] \ .
\ee
 $K$ is the hopping parameter and the matrix for the gauge-field link
 in the adjoint representation is defined as
\be  \label{eq09}
V_{rs,x\mu} \equiv V_{rs,x\mu}[U] \equiv
2 {\rm Tr}(U_{x\mu}^\dagger T_r U_{x\mu} T_s)
= V_{rs,x\mu}^* =V_{rs,x\mu}^{-1T} \ .
\ee
 The generators $T_r \equiv \half \lambda_r$ satisfy the usual
 normalization ${\rm Tr\,}(\lambda_r\lambda_s)=\half\delta_{rs}$.
 In case of SU(2) we have $T_r \equiv \half \tau_r$ with the isospin
 Pauli-matrices $\tau_r$.
 Starting from the Dirac fermion fields one can introduce two
 Dirac-Majorana fields $\Psi^{(1,2)}$ satisfying (\ref{eq02}):
\be  \label{eq10}
\Psi^{(1)} \equiv \frac{1}{\sqrt{2}} ( \psi + C\overline{\psi}^T)
\ , \hspace{2em}
\Psi^{(2)} \equiv \frac{i}{\sqrt{2}} (-\psi + C\overline{\psi}^T)
\ee
 and $S_f$ can be rewritten as
\be  \label{eq11}
S_f = \half\sum_{j=1}^2
\sum_{xu,yv} \overline{\Psi}^{(j)v}_y Q_{yv,xu} \Psi^{(j)u}_x \ .
\ee
 Using this, the fermionic path integral for Dirac fermions becomes
\be  \label{eq12}
\int [d\overline{\psi} d\psi] e^{-S_f} = 
\int [d\overline{\psi} d\psi] e^{-\overline{\psi} Q \psi} = \det Q
= \prod_{j=1}^2 \int [d\Psi^{(j)}] 
e^{ -\half\overline{\Psi}^{(j)}Q\Psi^{(j)} } \ .
\ee
 For Majorana fields the path integral involves only $[d\Psi^{(j)}]$,
 either with $j=1$ or $j=2$ hence, omitting the index $(j)$, we have
\be  \label{eq13}
\int [d\Psi] e^{ -\half\overline{\Psi} Q \Psi }
= \int [d\Psi] e^{ -\half\Psi^T CQ\Psi }
= {\rm Pf}(CQ) = {\rm Pf}(M) \ .
\ee
 Here the {\em Pfaffian} of the antisymmetric matrix $M \equiv CQ$
 is introduced.
 The Pfaffian can be defined for a general complex antisymmetric matrix
 $M_{\alpha\beta}=-M_{\beta\alpha}$ with an even number of dimensions
 ($1 \leq \alpha,\beta \leq 2N$) by a Grassmann integral as
\be  \label{eq14}
{\rm Pf}(M) \equiv
\int [d\phi] e^{-\half\phi_\alpha M_{\alpha\beta} \phi_\beta}
= \frac{1}{N!2^N} \epsilon_{\alpha_1\beta_1 \ldots \alpha_N\beta_N}
M_{\alpha_1\beta_1} \ldots M_{\alpha_N\beta_N} \ .
\ee
 Here, of course, $[d\phi] \equiv d\phi_{2N} \ldots d\phi_1$, and 
 $\epsilon$ is the totally antisymmetric unit tensor.
 It can be easily shown that
\be  \label{eq15}
\left[{\rm Pf}(M)\right]^2 = \det M \ .
\ee
 One way to prove this is to use $\det M = \det CQ = \det Q$ and
 eqs.~(\ref{eq12})-(\ref{eq13}).
 Besides the partition function in (\ref{eq12}), expectation values for
 Majorana fermions can also be similarly defined \cite{GLUINO,DGHV}.

 It is easy to show \cite{EDINBURGH} that the adjoint fermion matrix
 $Q$ has doubly degenerate real eigenvalues, therefore $\det Q$ is
 positive and ${\rm Pf}(M)$ is real.
 Omitting the sign of ${\rm Pf}(M)$ one obtains the effective gauge
 field action \cite{CURVEN}:
\be\label{eq16}
S_{CV} = \beta\sum_{pl} \left( 1-\half{\rm Tr\,}U_{pl} \right)
- \half\log\det Q[U] \ ,
\ee
 with the bare gauge coupling given by $\beta \equiv 2N_c/g^2$.
 The factor $\half$ in front of $\log\det Q$ tells that we effectively
 have a flavour number $N_f=\half$ of adjoint fermions.
 The omitted sign of the Pfaffian can be taken into account in the
 expectation values:
\be\label{eq17}
\langle A \rangle = \frac{\langle A\; {\rm sign\,Pf}(M)\rangle_{CV}}
{\langle {\rm sign\,Pf}(M)\rangle_{CV}} \ .
\ee
 This sign problem is very similar to the one in QCD with an odd
 number of quark flavours.

 The value of the Pfaffian, hence its sign, can be numerically
 determined by calculating an appropriate determinant \cite{BOULDER}.
 It turns out that in updating sequences with dynamical gauginos
 configurations with positive Pfaffian dominate.
 This is shown by explicit evaluation on $4^3 \cdot 8$ lattices.
 It is plausible that the sign changes, as a function of the valence
 hopping parameter, typically occur at higher values than the value of
 $K$ in the dynamical updating \cite{BOULDER}.
 Therefore, in the present work, we consider the effective gauge action
 in (\ref{eq16}) and neglect the sign of the Pfaffian.
 To take into account the sign is possible but numerically demanding,
 therefore we postpone it for future studies.

 Since the Monte Carlo calculations are done on finite lattices, one
 has to specify boundary conditions.
 In the three spatial directions we take periodic boundary conditions
 both for the gauge field and the gaugino.
 This implies that in the Hilbert space of states the supersymmetry is
 not broken by the boundary conditions.
 In the time direction we take periodic boundary conditions for bosons
 and antiperiodic ones for fermions, which is obtained if one writes
 traces in terms of Grassmann integrals.
 (The minus sign for fermions is the usual one associated with closed
 fermion loops.)
 Of course, boundary conditions do not influence the physical results
 in large volumes.
 For instance, we explicitly checked that the distribution of the
 gaugino condensate is not effected if in the time direction periodicity
 is assumed for the fermions, too (see below).
 Another interesting possibility would be to consider twisted boundary
 conditions \cite{THOOFT} which are useful in theoretical considerations
 about supersymmetry breaking \cite{WITTEN}.

%%%%%%%%%%%%%%%%%%%%%%%%%%%%%%%%%%%%%%%%%%%%%%%%%%%%%%%%%%%%%%%%%%%%%%%%
\section{Monte Carlo simulation}\label{sec3}
 The expected first order phase transition at zero gaugino mass should
 show up as a jump in the expectation value of the gaugino condensate
 (\ref{eq05}).
 The renormalized gaugino mass is obtained from the hopping parameter
 $K$ as
\be\label{eq18}
m_{R\tilde{g}} = \frac{Z_m(a\mu)}{2a} 
\left[ \frac{1}{K} - \frac{1}{K_0} \right]
\equiv  Z_m(a\mu) m_{0\tilde{g}} \ .
\ee
 Here $a$ denotes the lattice spacing, $\mu$ is the renormalization
 scale and $K_0=K_0(\beta)$ gives the $\beta$-dependent position of the
 phase transition, which is expected to approach $K_0=1/8$ in the
 continuum limit $\beta \to \infty$.
 The bare gaugino mass $m_{0\tilde{g}}$ is defined, as usual, by
 omitting the multiplicative renormalization factor $Z_m$.
 The renormalized gaugino condensate is also obtained by additive and
 multiplicative renormalizations:
\be\label{eq19}
\langle \bar{\Psi}_x \Psi_x \rangle_{R(\mu)} = Z(a\mu)
\left[ \langle \bar{\Psi}_x \Psi_x \rangle - b_0(a\mu) \right] \ .
\ee
 The renormalization factors $Z_m$ and $Z$ are expected to be of order
 ${\cal O}(1)$.
 The presence of the additive shift in the gaugino condensate
 $b_0(a\mu)$ implies that the value of its jump at $m_{R\tilde{g}}=0$
 is easier available than the value itself.

 A first order phase transition should show up on small to moderately
 large lattices as metastability expressed by a two-peak structure
 in the distribution of some {\em order parameter}, in our case the
 value of the gaugino condensate.
 By tuning the bare parameters in the action, in our case the hopping
 parameter $K$ for fixed gauge coupling $\beta$, one can achieve that
 the two peaks are equal (in height or area).
 This is the definition of the phase transition point in finite volumes.
 By increasing the volume the tunneling between the two ground states
 becomes less and less probable and at some point practically
 impossible.
 
 In our simulations, besides the distribution of the gaugino condensate,
 we also studied other quantities as the string tension or the masses of
 the lightest bound states.
 The first results have been published recently \cite{BOULDER,SPANDEREN}
 together with a first hint for the existence of a phase transition from
 a simulation at ($\beta=2.3,K=0.195$).
 In the present paper we keep the gauge coupling at $\beta=2.3$ and
 exploit the region around $K=0.195$.

 The Monte Carlo simulations are done by a two-step variant of the
 multi-bosonic algorithm \cite{LUSCHER} proposed in \cite{GLUINO}.
 We use polynomial approximations discussed in detail in \cite{POLYNOM}
 and correction procedures which are adapting some known methods from
 the literature \cite{KENKUT,FREJAN} to the present situation with
 $N_f=\half$ flavours.
 Our experience with this algorithm has been described already in
 previous publications \cite{DESYMUNSTER,BOULDER,SPANDEREN} and will
 be discussed in detail in a forthcoming paper \cite{LONG}.

%%%%%%%%%%%%%%%%%%%%%%%%%%%%%%%%%%%%%%%%%%%%%%%%%%%%%%%%%%%%%%%%%%%%%%%%
\begin{table}[ht]
\begin{center}
\parbox{15cm}{\caption{\label{tab01}\it
 Parameters of the numerical simulations on $6^3 \cdot 12$ lattice
 at $\beta=2.3$.
}}
\end{center}
\begin{center}
\begin{tabular}{| l | l | c | c | c | c | c | r | c | l | l |}
\hline
\multicolumn{1}{|c|}{$K$}  &
\multicolumn{1}{|c|}{$\epsilon$}  &
$\lambda$  & $n_1$  &  $n_2$  &  $n_3$  &  $n_4$  &
\multicolumn{1}{|c|}{updates} &  $A_{nc}$  &
\multicolumn{1}{|c|}{$\tau_{\rm plaq}$}  &
\multicolumn{1}{|c|}{$C_\rho^{(240)}$}   \\
\hline\hline
0.19    &  0.0005   &  3.6  &  20  &  112  &  150  &  400  &
1487360  &  0.888   &  214(9)  &  0.136(42)  \\
\hline
0.1925  &  0.0001   &  3.7  &  22  &  132  &  180  &  400  &
3655680  &  0.889   &  220(7)  &  0.220(36)  \\
\hline
0.195   &  0.00001  &  3.7  &  24  &  200  &  300  &  400  &
460800  &   0.892   &  256(15) &  0.063(38)  \\
\hline
0.195$^\ast$   &  0.00003  &  3.7  &  22  &  66  &  102  &  400  &
1224000   &  0.823  &
\multicolumn{1}{|c|}{-}  &  \multicolumn{1}{|c|}{-} \\
\hline
0.196   &  0.00001  &  3.7  &  24  &  200  &  300  &  400  &
952320  &  0.889    &  321(26)  &  0.180(32)  \\
\hline
0.1975  &  0.000001 &  3.8  &  30  &  300  &  400  &  500  &
506880  &  0.926    &  295(17)  &  0.367(31)  \\
\hline
0.2     &  0.000001 &  3.9  &  30  &  300  &  400  &  500  &
599040  &  0.925    &  317(16)  &  0.424(26)  \\
\hline\hline
\end{tabular}
\end{center}
\end{table}
%%%%%%%%%%%%%%%%%%%%%%%%%%%%%%%%%%%%%%%%%%%%%%%%%%%%%%%%%%%%%%%%%%%%%%%%
 The parameters of the numerical simulations on $6^3 \cdot 12$ lattice
 at $\beta=2.3$ are summarized in table~\ref{tab01}.
 The run with an asterisk had periodic boundary conditions for the
 gaugino in the time direction, the rest antiperiodic.
 $K$ is the hopping parameter and $[\epsilon,\lambda]$ is the interval
 of approximation for the first three polynomials of orders
 $n_{1,2,3}$, respectively.
 The fourth polynomial of order $n_4$ is defined on $[0,\lambda]$.
 In the eighth column the number of performed updating cycles is given.
 The ninth column contains the acceptance rate in the noisy correction
 step $A_{nc}$, the tenth column gives the exponential autocorrelation
 length for plaquettes $\tau_{\rm plaq}$ observed in the range of about
 100 updating steps.
 The integrated autocorrelation is roughly a factor four higher, with
 large errors: for instance at $K=0.1925$
 $\tau_{\rm plaq}^{\rm int} \simeq 900 \pm 300$.
 The last column contains the value of the autocorrelation function
 of the gaugino condensate at a distance 240, where the measurements
 were performed.

\begin{figure}[tb]
\begin{flushleft}
\epsfig{file=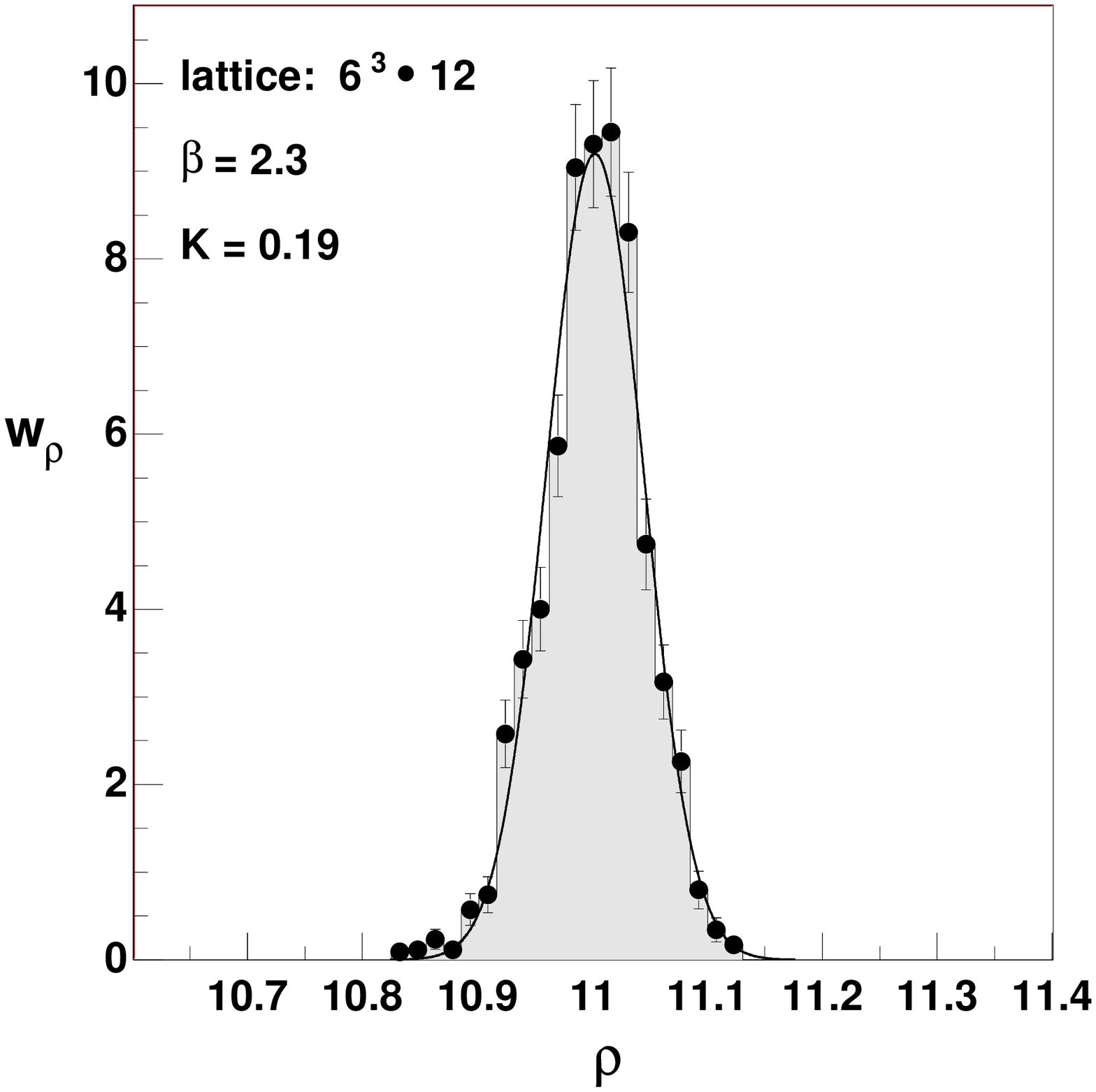,
        width=75mm,height=65mm,
       %bbllx=77pt,bblly=404pt,bburx=566pt,bbury=686pt,
        angle=0}
\end{flushleft}
\vspace*{-74mm}
\begin{flushright}
\epsfig{file=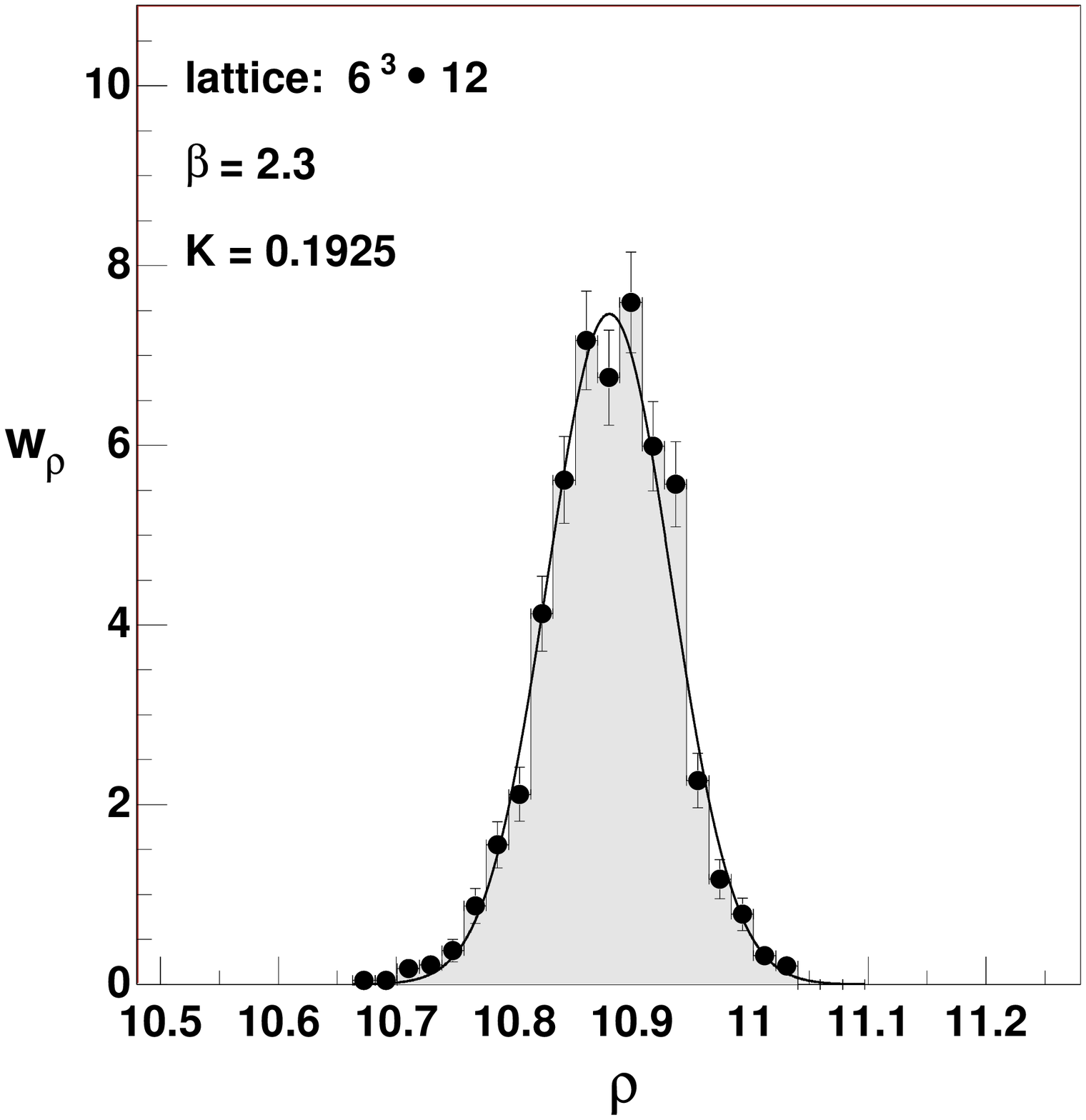,
        width=75mm,height=65mm,
       %bbllx=77pt,bblly=404pt,bburx=566pt,bbury=686pt,
        angle=0}
\end{flushright}
\begin{flushleft}
\epsfig{file=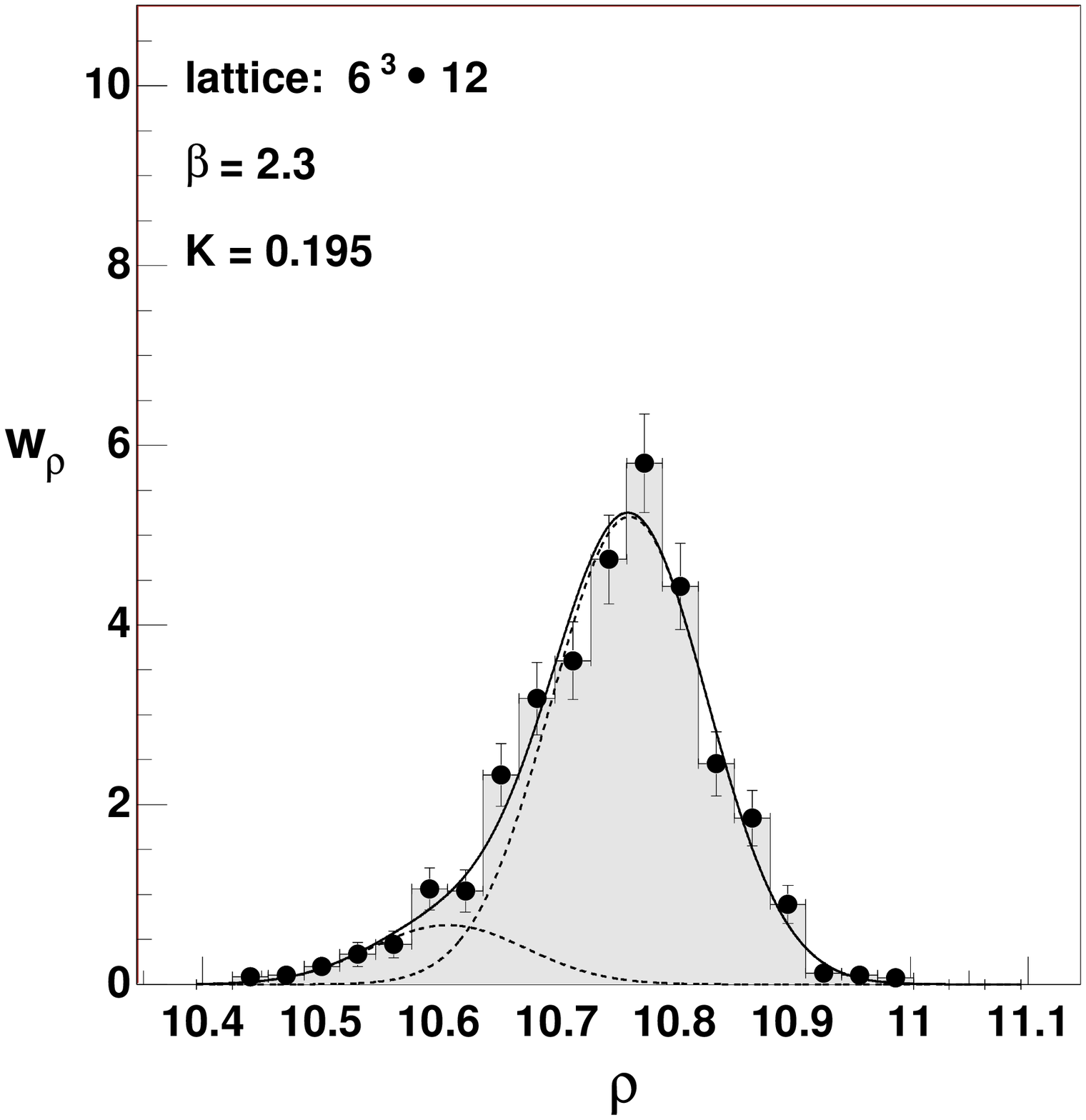,
        width=75mm,height=65mm,
       %bbllx=77pt,bblly=404pt,bburx=566pt,bbury=686pt,
        angle=0}
\end{flushleft}
\vspace*{-74mm}
\begin{flushright}
\epsfig{file=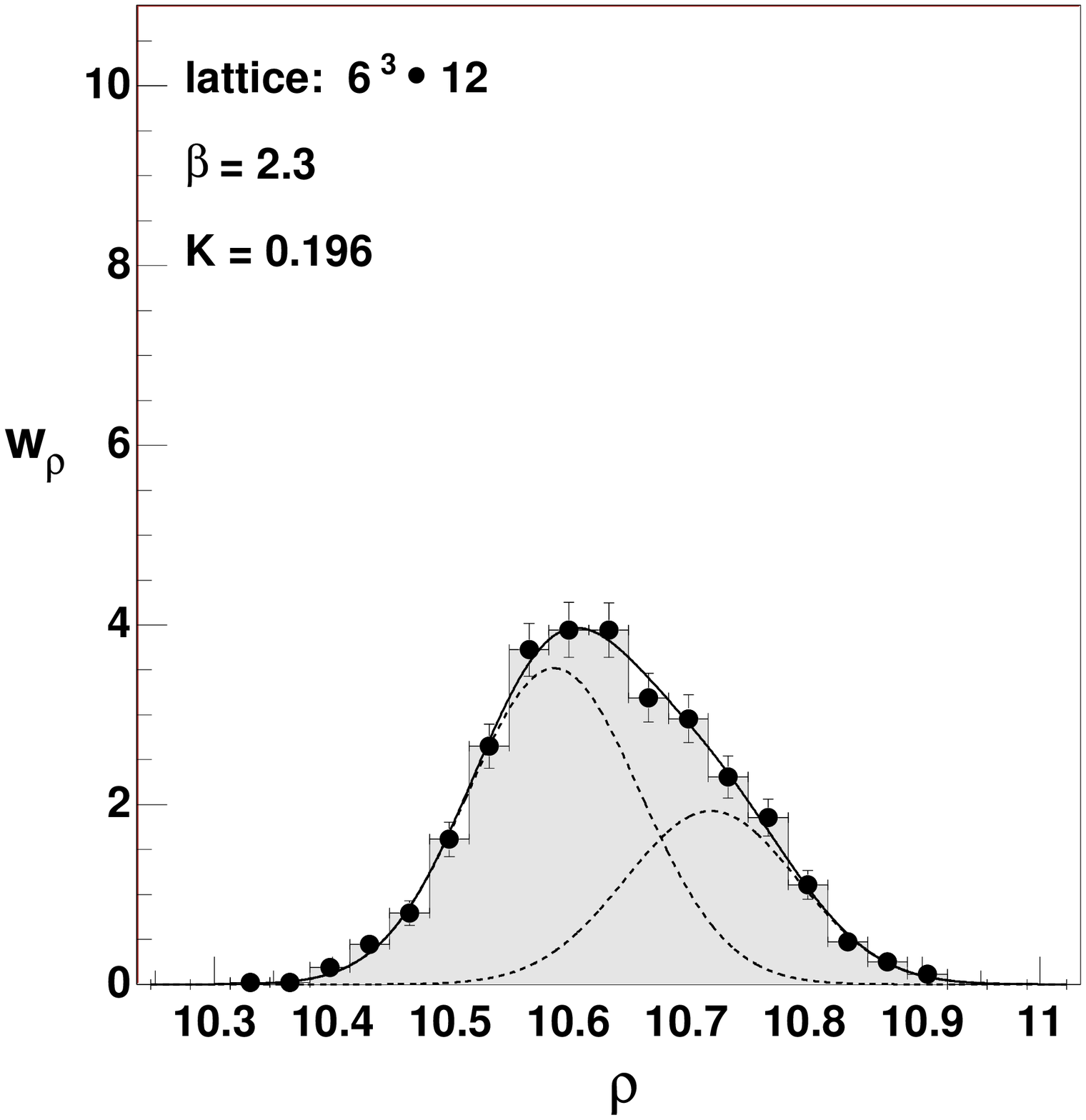,
        width=75mm,height=65mm,
       %bbllx=77pt,bblly=404pt,bburx=566pt,bbury=686pt,
        angle=0}
\end{flushright}
\begin{flushleft}
\epsfig{file=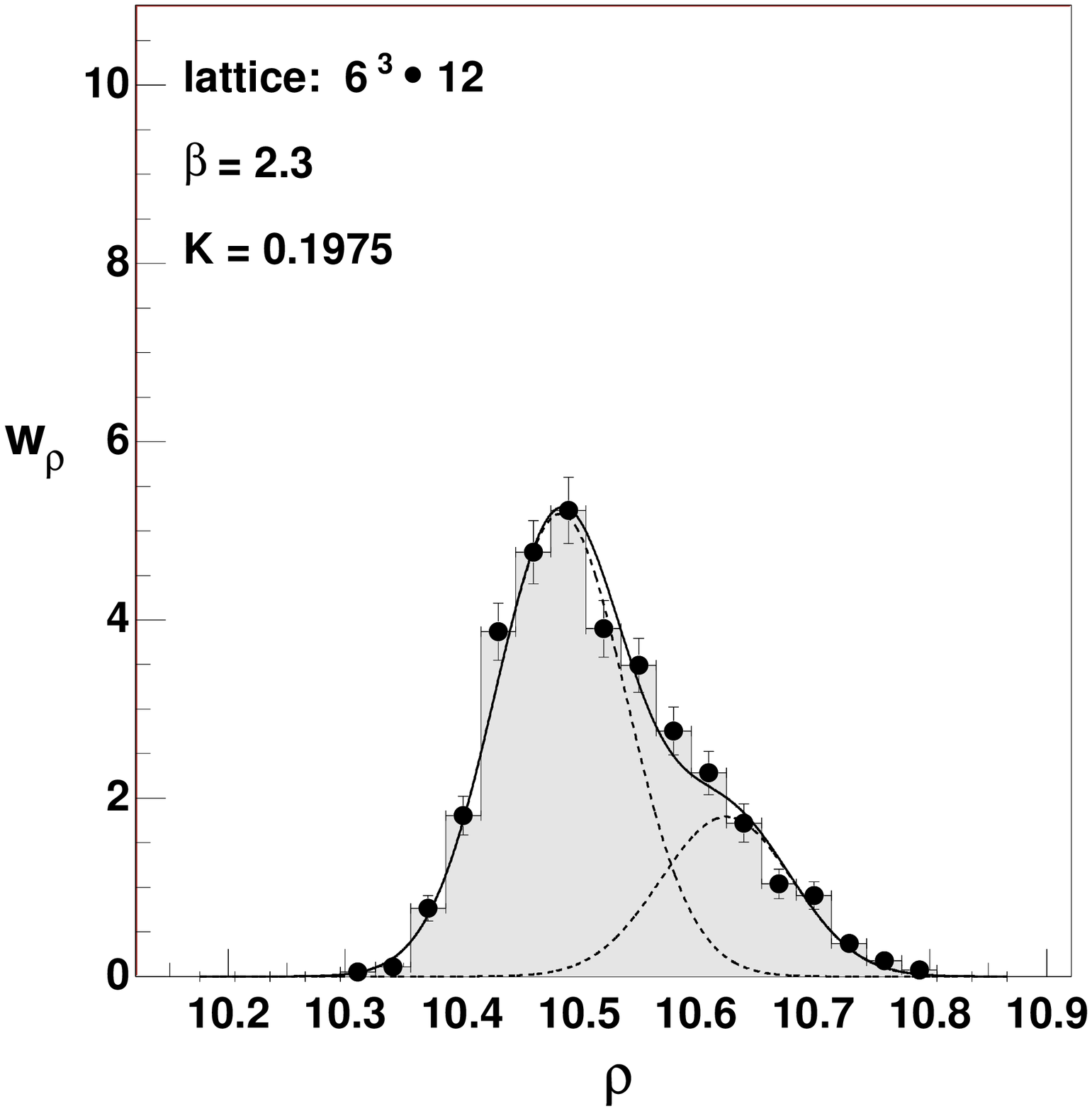,
        width=75mm,height=65mm,
       %bbllx=77pt,bblly=404pt,bburx=566pt,bbury=686pt,
        angle=0}
\end{flushleft}
\vspace*{-74mm}
\begin{flushright}
\epsfig{file=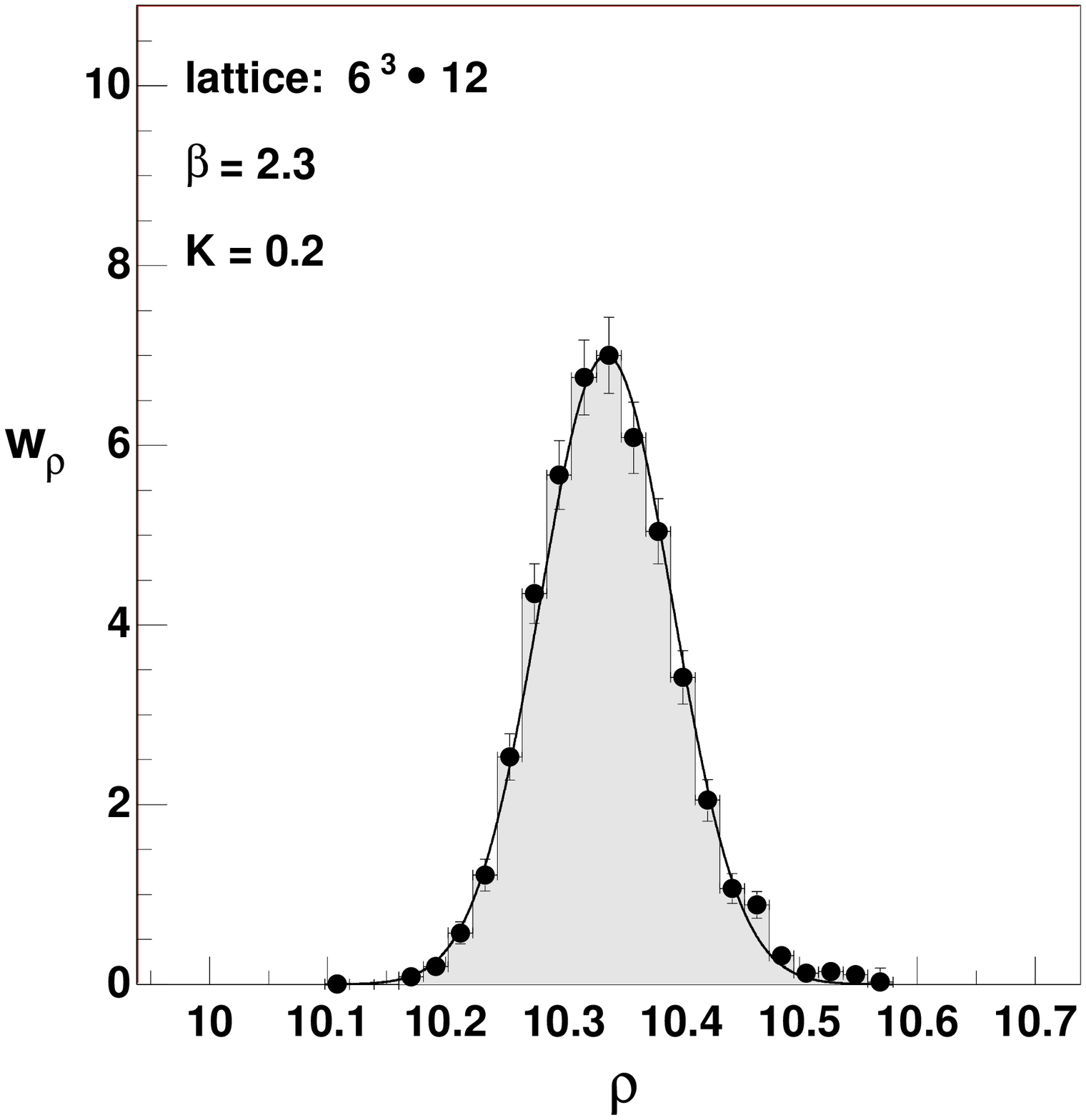,
        width=75mm,height=65mm,
       %bbllx=77pt,bblly=404pt,bburx=566pt,bbury=686pt,
        angle=0}
\end{flushright}
\vspace*{-10mm}
\begin{center}
\parbox{16cm}{\caption{\label{fig01}
 The probability distributions of the gaugino condensate for different
 hopping parameters at $\beta=2.3$ on $6^3 \cdot 12$ lattice.
 The dased lines show the Gaussian components.
}}
\end{center}
\end{figure}

 The order parameter of the supersymmetry phase transition at zero
 gaugino mass is the value of the gaugino condensate
\be\label{eq20}
\rho \equiv 
\frac{1}{\Omega}\sum_x \left(\bar{\Psi}_x \Psi_x\right) \ .
\ee
 The normalization is provided by the number of lattice points $\Omega$.
 We determined the value of $\rho$ on a gauge configuration by
 stochastic estimators
\be\label{eq21}
\frac{1}{N_\eta}\sum_{i=1}^{N_\eta}
\sum_{xy}\left(\bar{\eta}_{y,i} Q^{-1}_{yx}\eta_{x,i}\right)
\ee
 on normalized Gaussian random vectors $\eta_{x,i}$.
 In practice $N_\eta=25$ works fine.
 Outside the phase transition region the observed distribution of $\rho$
 can be fitted well by a single Gaussian, but in the transition region
 a reasonably good fit can only be obtained with two Gaussians
 (see figure~\ref{fig01}).
 The fit parameters of the distributions ($i=1$ or $i=1,2$)
\be\label{eq22}
\frac{p_i}{\sigma_i\sqrt{2\pi}}
\exp\left\{ -\frac{(\rho-\mu_i)^2}{2\sigma_i^2} \right\}
\ee
 and the $\chi^2$ values per degrees of freedom are given in
 table~\ref{tab02}.
 The normalization is such that $p_1+p_2=1$.
 Exact supersymmetry would imply that the widths of the two Gaussians
 are equal.
 This relation is broken by the lattice regularization and by the 
 non-zero gaugino mass away from the phase transition point.
 In order to keep the number of fit parameters small we neglect this
 small symmetry breaking and the fits are done under the assumption
 $\sigma_1=\sigma_2\equiv\sigma$.
 The statistical errors of the fit parameters are determined by
 jack-knifing 64  statistically independent parallel runs.

%%%%%%%%%%%%%%%%%%%%%%%%%%%%%%%%%%%%%%%%%%%%%%%%%%%%%%%%%%%%%%%%%%%%%%%%
\begin{table}[ht]
\begin{center}
\parbox{15cm}{\caption{\label{tab02}\it
 Fit parameters of the order parameter distributions corresponding to
 the runs in table~\protect\ref{tab01}.
 The statistical errors in last digits are given in parentheses.
}}
\end{center}
\begin{center}
\begin{tabular}{| l | l | l | l | l | l |}
\hline
\multicolumn{1}{|c|}{$K$}              &
\multicolumn{1}{|c|}{$p_1$}            &  
\multicolumn{1}{|c|}{$\mu_1$}          &  
\multicolumn{1}{|c|}{$\mu_2$}          & 
\multicolumn{1}{|c|}{$\sigma$}         &
\multicolumn{1}{|c|}{$\chi^2/d.o.f.$}  \\
\hline\hline
0.19   &   1.0   &  11.0023(26) & \multicolumn{1}{|c|}{-} & 
0.0423(16) & 27.9/20 \\
\hline
0.1925 &   1.0   &  10.8807(30) & \multicolumn{1}{|c|}{-} & 
0.0524(17) & 25.9/20 \\
\hline
0.195  & 0.89(7) &  10.762(30)  &  10.608(30)  &  0.066(7)  & 16.5/18 \\
\hline
0.195$^\ast$ & 0.83(6) & 10.78(3) &  10.60(3)  &  0.055(7)  & 16.3/18 \\
\hline
0.196  & 0.35(7) &  10.722(11)  &  10.588(11)  &  0.073(3)  &  5.7/18 \\
\hline
0.1975 & 0.26(5) &  10.626(17)  &  10.484(17)  &  0.056(4)  & 19.5/18 \\
\hline
0.2    &   0.0   & \multicolumn{1}{|c|}{-} &
10.3363(37) & 0.0562(18) & 21.4/20 \\
\hline\hline
\end{tabular}
\end{center}
\end{table}
%%%%%%%%%%%%%%%%%%%%%%%%%%%%%%%%%%%%%%%%%%%%%%%%%%%%%%%%%%%%%%%%%%%%%%%%

 As figure~\ref{fig01} and table~\ref{tab02} show, in the region
 $0.195 \leq K \leq 0.1975$ the distribution of the gaugino condensate
 can only be fitted well by two Gaussians.
 Comparing the two runs at $K=0.195$ with antiperiodic, respectively,
 periodic boundary conditions in the time direction, one can see that
 the different boundary conditions do not have a sizeable effect
 on the distributions, as remarked before.
 For increasing $K$ (decreasing bare gaugino mass) the weights shift
 from the Gaussian at larger $\rho$ to the one with smaller $\rho$, as
 expected.
 The two Gaussians represent the contributions of the two phases on this
 lattice.
 The position of the phase transition on the $6^3 \cdot 12$ lattice
 is at $K_0=0.1955 \pm 0.0005$.
 The jump of the order parameter is
 $\Delta\rho \equiv \mu_1 - \mu_2 \simeq 0.15$.

 The two-phase structure can also be searched for in pure gauge field
 variables as the plaquette or longer Wilson loops.
 It turns out that the distributions of Wilson loops are rather
 insensitive.
 They can be well described by single Gaussians with almost constant
 variance in the whole range $0.19 \leq K \leq 0.2$ (see, for instance,
 table~\ref{tab03}).
 This speaks against the appearance of a third chirally symmetric phase
 \cite{EVAHSUSCH}, which has been suggested in \cite{THIRD}.
%%%%%%%%%%%%%%%%%%%%%%%%%%%%%%%%%%%%%%%%%%%%%%%%%%%%%%%%%%%%%%%%%%%%%%%%
\begin{table}[h]
\begin{center}
\parbox{15cm}{\caption{\label{tab03}\it
 Fit parameters of the plaquette distributions on $6^3 \cdot 12$ lattice
 at $\beta=2.3$ for different hopping parameters.
 The statistical errors in last digits are given in parentheses.
}}
\end{center}
\begin{center}
\begin{tabular}{| l | l | l | l | l | l |}
\hline
\multicolumn{1}{|c|}{$K$}              &
\multicolumn{1}{|c|}{$p_1$}            &  
\multicolumn{1}{|c|}{$\mu_1$}          &
\multicolumn{1}{|c|}{$\sigma_1$}       &
\multicolumn{1}{|c|}{$\chi^2/d.o.f.$}  \\
\hline\hline
0.19   &  0.974(25)  &  0.63165(8)  &  0.00425(13)  &  0.89/47  \\
\hline
0.1925 &  1.014(27)  &  0.63511(8)  &  0.00461(15)  &  0.74/47  \\
\hline
0.195  &  0.997(59)  &  0.63811(19) &  0.00481(35)  &  2.58/47  \\
\hline
0.196  &  1.059(63)  &  0.64182(22) &  0.00518(36)  &  1.77/47  \\
\hline
0.1975 &  0.987(54)  &  0.64452(18) &  0.00444(30)  &  2.26/47  \\
\hline
0.2    &  1.018(44)  &  0.64846(13) &  0.00424(22)  &  2.00/47  \\
\hline\hline
\end{tabular}
\end{center}
\end{table}
%%%%%%%%%%%%%%%%%%%%%%%%%%%%%%%%%%%%%%%%%%%%%%%%%%%%%%%%%%%%%%%%%%%%%%%%

%%%%%%%%%%%%%%%%%%%%%%%%%%%%%%%%%%%%%%%%%%%%%%%%%%%%%%%%%%%%%%%%%%%%%%%%
%%%%%%%%%%%%%%%%%%%%%%%%%%%%%%%%%%%%%%%%%%%%%%%%%%%%%%%%%%%%%%%%%%%%%%%%

%%%%%%%%%%%%%%%%%%%%%%%%%%%%%%%%%%%%%%%%%%%%%%%%%%%%%%%%%%%%%%%%%%%%%%%%
\newpage
\section{Summary and discussion}\label{sec4}
 The observed dependence of the distribution of the gaugino condensate
 on the gaugino mass $m_{\tilde{g}}$ near $m_{\tilde{g}}=0$ is
 consistent with a typical behaviour characteristic of a first order
 phase transition between two phases (see figure~\ref{fig01} and
 table~\ref{tab02}).
 Our lattice volume ($L^3 \cdot T=6^3 \cdot 12$) is, however, still
 not very large in physical units, therefore the expected two-peak
 structure is not yet well developed.
 For instance, at $K=0.1925$ we have $LM_{gg}^{0^+} \simeq 3.6$, with
 the smallest glueball mass $M_{gg}^{0^+}$ \cite{BOULDER,LONG}.
 In fact, a behaviour corresponding to a true first order phase
 transition can only be established in a detailed study of the volume
 dependence, which we postpone for future work.
 Therefore, the present observations are also consistent with a rapid
 cross-over at finite lattice spacings, approaching to a first order
 phase transition in the continuum limit $\beta\to\infty$.
 On our $6^3 \cdot 12$ lattice for $\beta=2.3$ the phase transition
 (or cross-over) is at $K_0=0.1955 \pm 0.0005$.
 The jump of the gaugino condensate in lattice units is
 $\Delta\rho \simeq 0.15$.

 A rather positive aspect of our Monte Carlo simulations is the 
 ability of the two-step multi-bosonic algorithm \cite{GLUINO} to
 cope with the difficult situation at small dynamical fermion mass in
 the environment of metastability of phases.

 In the numerical simulations we considered up to now only the
 unrenormalized gaugino mass and gaugino condensate.
 The transformation to the corresponding renormalized quantities
 defined in eqs.~(\ref{eq18})-(\ref{eq19}) will, however, not change
 the qualitative behaviour, because the multiplicative renormalization
 constants are expected to be of ${\cal O}(1)$.
 One has to note that in the exploited range the bare gaugino masses
 $m_{0\tilde{g}}$ are small compared to the lightest bound state masses.
 With $K_0=0.1955$ at $K=0.1925$ we have
 $m_{0\tilde{g}}/M_{gg}^{0^+} \simeq 0.07$.
 Similarly to QCD, it is expected that the mass gap in the spectrum
 is of the same order of magnitude as the scale parameter for the
 asymptotically free coupling $\Lambda$.
 As the preliminary results on the bound state masses show
 \cite{BOULDER,LONG}, at $K=0.1925$ we already have an approximate
 degeneracy of the states which are expected to form the lowest chiral
 supermultiplet.

 Besides the volume dependence, another interesting question is the
 development of the phase transition signal towards the continuum limit
 at $\beta=\infty$.
 In fact, the arguments in the introduction (at
 eqs.~(\ref{eq01})-(\ref{eq06})) for the spontaneous chiral symmetry
 breaking $Z_{2N_c} \to Z_2$ refer to the continuum limit.
 The present numerical evidence shows that the discrete chiral
 symmetry breaking is manifested at non-zero lattice spacing in feasible
 numerical simulations and can be investigated by well established
 methods.

 {\bf Acknowledgements:} It is a pleasure to thank Gernot M\"unster
 for helpful discussions.
 The numerical simulations presented here have been performed on the
 CRAY-T3E-512 computer at HLRZ J\"ulich.
 We thank HLRZ and the staff at ZAM for their kind support.

%%%%%%%%%%%%%%%%%%%%%%%%%%%%%%%%%%%%%%%%%%%%%%%%%%%%%%%%%%%%%%%%%%%%%%%%

\end{document}